\documentclass{article}
\begin{document}
\begin{center}
{\bf \Large CNOT operator and its similar matrices in quantum computation}\\
\bigskip
Z. Sazonova$^{a}$, R. Singh$^{b}$\\
\bigskip
$^{a}$Physics Department, Moscow Automobile $\&$ Road Construction 
Institute (Technical University), 64, Leningradskii prospect, 
Moscow 125829, Russia\\
\bigskip
$^{b}$General Physics Institute of Russian Academy of Sciences, 38, Vavilov 
street, Moscow 117942, Russia, email: ranjit@dataforce.net
\end{center}
\begin{abstract}
We present the theoretical result, which is based on the linear algebra theory 
(similar operators). The obtained theoretical results optimize the 
experimental technique to construct quantum computer e.g., reduces the 
number of steps to perform the logical CNOT (XOR) operation. The present 
theoretical technique can also be generalized to the other operators in 
in quantum computation and information theory. 
\end{abstract}
\section{Introduction}
The linear algebra, which is the nucleus of quantum theory, has many 
interesting properties in the theoretical predictions of physical processes.
Tensor product [3], which is very widely, used in quantum theory plays an 
important role not only for obtaining the higher dimensional Hilbert space 
but also to optimize the experimental technique. Since, the tensor product 
is non-commutative in nature i.e., it can not be applied freely to any 
operator without any knowledge of the constituents of the physical system.  
One very important property of tensor product is similar matrices [3],
which have two optimization properties: reduces the number of pulses to 
realize some operator in physical experiment, and provides different choices of 
pulse propagations along different directions of axes. These two properties 
will be discussed in section 2 with concrete physical realization of CNOT 
operator and how we can optimize its realization.

\section{CNOT gates and CNOT's similar matrices in quantum computation}
CNOT gates or similar matrices play an important role in the construction 
of quantum computer and computation. All the complex quantum algorithms
are based on the combination of NOT and CNOT logical gates or matrices 
in quantum computation. Here, we will not discuss the NOT logical gate or
matrix, which is very simple negation operation. We will go in deep the
mathematical and physical nature of CNOT logical gates. The other complex
gates can be constructed on the basis of CNOT logical gates. Moreover, 
the construction of CNOT matrices plays an important role in quantum 
computation. Here, we will show that the CNOT matrices, which were used 
by [1,2] can be optimized by using the similar matrices i.e., we can find 
different CNOT matrices with the same mathematical properties but some 
different experimental or physical realization. For better insight to 
understand the physical nature of similar matrices, we will discuss in 
the whole article the system of two spins $\hat{\sigma}_{1}=1/2$ and 
$\hat{\sigma}_{2}=1/2$ with slightly different resonance frequencies 
$\omega_{1}$ and $\omega_{2}$ and having scalar coupling $\omega_{12}$. 
The Hamiltonain of the two spins aligned along the $z$-axis with the 
constant magnetic filed
\begin{eqnarray}
\hat{H}=\hbar\omega_{1}\hat{\sigma}_{1z}\otimes{\hat{e}_{2}}+\hbar\omega_{2}\hat{e}_{1}\otimes{\hat{\sigma}_{2z}}+\hbar\omega_{12}\hat{\sigma}_{1z}\otimes{\hat{\sigma}_{2z}}.
\end{eqnarray}
Where $\hat{e}_{i\in{(1,2)}}$ - identity matrix with dimensions $2\times{2}$ and $\hbar$ -
is Plank's constant. 

\subsection{Physical differences between the CNOT matrices obtained by Gershenfeld and Cory}
The CNOT matrix used by [2] contains additional spin $\hat{\sigma}_{1}$ 
rotation around the $z$-axis, which is unnecessary and which complicate 
the experiment realization of CNOT as compare to the CNOT matrix used by
Cory [1]. In the mathematical point of view, CNOT matrices obtained by 
Gershenfeld and Cory have different mathematical properties i.e., 
CNOT matrices (2) and (4,5) are not similar matrices due to the 
additional rotation of spin $\hat{\sigma}_{z1}$ around $z$-axis in (3) as
compare to the (6,7).

Moreover, additional rotation takes more time to fulfill the CNOT 
operation, which slows down the computation of quantum algorithms.
The CNOT matrix for Hamiltonain (1) obtained by Gershenfeld [2] is
\begin{eqnarray}
C_{g}=\sqrt{-i}
\left(
\begin{array}{cccc}
1 & 0 & 0 & 0\\
0 & 1 & 0 & 0\\
0 & 0 & 0 & 1\\
0 & 0 & 1 & 0
\end{array}
\right).
\end{eqnarray}
The $C_{g}$ matrix was obtained by the following pulse sequences
\begin{eqnarray}
C_{g}=R_{y2}(-\pi/4)R_{z1}(-\pi/4)R_{z2}(-\pi/4)R_{z12}(\pi/4)R_{y2}(\pi/4) \nonumber \\
=e^{-i\frac{\pi}{4}\hat{e}_{1}\otimes{\hat{\sigma}_{y2}}}e^{-i\frac{\pi}{4}\hat{\sigma}_{z1}\otimes{\hat{e}_{2}}}e^{-i\frac{\pi}{4}\hat{e}_{1}\otimes{\hat{\sigma}_{z2}}}e^{i\frac{\pi}{4}\hat{\sigma}_{z1}\otimes{\hat{\sigma}_{z2}}}e^{i\frac{\pi}{4}\hat{e}_{1}\otimes{\hat{\sigma}_{z2}}}.
\end{eqnarray}
Where $R$ - is rotation matrix around the different axes along with the 
different angles around different axes. The physical interpretation
of matrix $R$ is just the rotation of our physical reference system or quantum
tomography [4].

The two CNOT matrices for Hamiltonain (1) obtained by Cory [1] are
\begin{eqnarray}
C_{c1}=
\left(
\begin{array}{cccc}
1 & 0 & 0 & 0\\
0 & 1 & 0 & 0\\
0 & 0 & 0 & 1\\
0 & 0 &-1 & 0
\end{array}
\right).\\
C_{c2}=
\left(
\begin{array}{cccc}
1 & 0 & 0 & 0\\
0 & 0 & 0 & 1\\
0 & 0 & 1 & 0\\
0 &-1 & 0 & 0
\end{array}
\right).
\end{eqnarray}
The $C_{c1}$ and $C_{c2}$ matrices were obtained with the following 
pulse sequences 
\begin{eqnarray}
C_{c1}=R_{x2}(-\pi/4)R_{z2}(-\pi/4)R_{z12}(\pi/4)R_{x2}(\pi/4) \nonumber \\
=e^{-i\frac{\pi}{4}\hat{e}_{1}\otimes{\hat{\sigma}_{x2}}}e^{-i\frac{\pi}{4}\hat{e}_{1}\otimes{\hat{\sigma}_{z2}}}e^{i\frac{\pi}{4}\hat{\sigma}_{z1}\otimes{\hat{\sigma}_{z2}}}e^{i\frac{\pi}{4}\hat{e}_{1}\otimes{\hat{\sigma}_{x2}}}.\\
C_{c2}=R_{x1}(-\pi/4)R_{z1}(-\pi/4)R_{z12}(\pi/4)R_{x1}(\pi/4) \nonumber \\
=e^{-i\frac{\pi}{4}\hat{\sigma}_{x1}\otimes{\hat{e}_{2}}}e^{-i\frac{\pi}{4}\hat{\sigma}_{z1}\otimes{\hat{e}_{2}}}e^{i\frac{\pi}{4}\hat{\sigma}_{z1}\otimes{\hat{\sigma}_{z2}}}e^{i\frac{\pi}{4}\hat{\sigma}_{x1}\otimes{\hat{e}_{2}}}.
\end{eqnarray}

\subsection{Mathematical differences between the CNOT matrices obtained by 
Gershenfeld and Cory}
Since, the Hamiltonian of the CNOT matrices (2) and (4,5) are the same but 
they are not similar matrices. To check the mathematical nature of 
(2) and (4,5), first of all we will write six mathematical properties of
similar square matrices $A$ and $B$ of dimensions $4\times{4}$ on the 
Hilbert space $F$.
\begin{itemize} 
\item{Determinant of $A$ is equal to determinant of $B$.}
\item{Trace of $A$ is equal to trace of $B$.}
\item{If $A$ and $B$ are nonsingular than $A^{-1}$ and $B^{-1}$ are also
similar matrices.}
\item{$A$ and $B$ are similar matrices, if there exist nonsingular matrix
$P$ such that $B=P^{-1}AP$ or $PBP^{-1}=A$.}
\item{Matrices $A$ and $B$ have the same eigenvalues.}
\item{$PB_{evecs}=A_{evecs}$, where $B_{evecs}$ and $A_{evecs}$ are the 
eigenvectors of the matrices $B$ and $A$.}
\end{itemize}

Here, we will not proof the properties of similar matrices. The proof of 
above six properties are very simple and can be found in the course of
linear algebra [5].

By applying six properties of similar matrices to the CNOT matrices (2),
(4,5), we will find that that the properties second, fourth, fifth 
and sixth are not satisfied between CNOT matrices (2) and (4,5). So, the 
matrices (2) and (4,5) are not similar.

\subsection{Physical interpretation of similar matrices}
The similar matrices are very important part of the linear algebra in quantum
computation. For example, by finding all the similar matrices, we can get all
the CNOT matrices or operators, which will reduce our mathematical and
physical realization of quantum computation. If we apply all the six 
properties of similar matrices to CNOT matrices (4) and (5), we will see 
that matrices (4) and (5) are similar matrices.

Now, if we apply operators $C_{c1}$ and $C_{c2}$ to the state $|\phi>=a|\uparrow,\uparrow>+b|\uparrow,\downarrow>+c|\downarrow,\uparrow>+d|\downarrow,\downarrow>$,
of Hamiltonian (1). We will obtain
\begin{eqnarray}
|\phi_{1}>=
C_{c1}|\phi>=
\left(
\begin{array}{cccc}
1 & 0 & 0 & 0\\
0 & 1 & 0 & 0\\
0 & 0 & 0 & 1\\
0 & 0 &-1 & 0
\end{array}
\right)
\left(
\begin{array}{c}
a(\uparrow,\uparrow)\\
b(\uparrow,\downarrow)\\
c(\downarrow,\uparrow)\\
d(\downarrow,\downarrow)
\end{array}
\right)
=
\left(
\begin{array}{c}
a(\uparrow,\uparrow)\\
b(\uparrow,\downarrow)\\
d(\downarrow,\downarrow)\\
-c(\downarrow,\uparrow)
\end{array}
\right).
\end{eqnarray}

\begin{eqnarray}
|\phi_{2}>=
C_{c2}|\phi>=
\left(
\begin{array}{cccc}
1 & 0 & 0 & 0\\
0 & 0 & 0 & 1\\
0 & 0 & 1 & 0\\
0 &-1 & 0 & 0
\end{array}
\right)
\left(
\begin{array}{c}
a(\uparrow,\uparrow)\\
b(\uparrow,\downarrow)\\
c(\downarrow,\uparrow)\\
d(\downarrow,\downarrow)
\end{array}
\right)
=
\left(
\begin{array}{c}
a(\uparrow,\uparrow)\\
d(\downarrow,\downarrow)\\
c(\downarrow,\uparrow)\\
-b(\uparrow,\downarrow)
\end{array}
\right).
\end{eqnarray}
the state $|\phi_{1,2}>$, which is obtained by operating $C_{c1}$ and $C_{c2}$,
which gives us simple picture of the state beofre and after the CNOT operation.
In the experiment of nuclear magnetic resonance, the CNOT matrices (4,5) 
are called Pound-Overhauser operators [1] i.e., transformation of spin 
polarization from one spin to other spin.

How can we find all the CNOT matrices or similar matrices of (4,6)? Is 
there any method to get all the similar matrices? One of the main purpose 
of this paper is to find answer to these questions.

We are not interested in finding similar matrices of CNOT matrix (2) 
because it has an additional rotation, which complicates CNOT operation and
which is not required for the CNOT operation. So, we are interested in 
finding the similar matrices of (4,5), which minimize CNOT operation in 
quantum computation. 

\subsection{How we can find CNOT similar matrices and how many are they?}
As it is seen from the (8) and (9) that the state $|\uparrow,\uparrow>$ in 
$|\phi>$ is not used by the CNOT matrix. It means, we have still more options 
to use the possibility to get other CNOT matrices or similar matrices. This
can be done by considering the six similar matrix properties.
\begin{eqnarray}
C_{c11}=R_{y2}(-\pi/4)R_{z12}(\pi/4)R_{z2}(-\pi/4)R_{y2}(\pi/4) \nonumber \\
=e^{-i\frac{\pi}{4}\hat{e}_{1}\otimes{\hat{\sigma}_{y2}}}e^{i\frac{\pi}{4}\hat{\sigma}_{z1}\otimes{\hat{\sigma}_{z2}}}e^{-i\frac{\pi}{4}\hat{e}_{1}\otimes{\hat{\sigma}_{z2}}}e^{i\frac{\pi}{4}\hat{e}_{1}\otimes{\hat{\sigma}_{y2}}} \nonumber \\
=\left(
\begin{array}{cccc}
1 & 0 & 0 & 0\\
0 & 1 & 0 & 0\\
0 & 0 & 0 & -i\\
0 & 0 & -i & 0
\end{array}
\right).\\
C_{c22}=
R_{y1}(-\pi/4)R_{z12}(\pi/4)R_{z1}(-\pi/4)R_{y1}(\pi/4) \nonumber \\
=e^{-i\frac{\pi}{4}{\hat{\sigma}_{y1}}\otimes{\hat{e}_{2}}}e^{i\frac{\pi}{4}\hat{\sigma}_{z1}\otimes{\hat{\sigma}_{z2}}}e^{-i\frac{\pi}{4}\hat{\sigma}_{z1}\otimes{\hat{e}_{2}}}e^{i\frac{\pi}{4}\hat{\sigma}_{y1}\otimes{\hat{e}_{2}}} \nonumber \\
=\left(
\begin{array}{cccc}
1 & 0 & 0 & 0\\
0 & 0 & 0 & -i\\
0 & 0 & 1 & 0\\
0 & -i & 0 & 0
\end{array}
\right).\\
C_{c31}=
R_{y1}(-\pi/4)R_{z12}(-\pi/4)R_{z1}(-\pi/4)R_{y1}(\pi/4) \nonumber \\
=e^{-i\frac{\pi}{4}{\hat{\sigma}_{y1}}\otimes{\hat{e}_{2}}}e^{-i\frac{\pi}{4}\hat{\sigma}_{z1}\otimes{\hat{\sigma}_{z2}}}e^{-i\frac{\pi}{4}\hat{\sigma}_{z1}\otimes{\hat{e}_{2}}}e^{i\frac{\pi}{4}\hat{\sigma}_{y1}\otimes{\hat{e}_{2}}} \nonumber \\
=\left(
\begin{array}{cccc}
0 & 0 & -i & 0\\
0 & 1 & 0 & 0\\
-i & 0 & 0 & 0\\
0 & 0 & 0 & 1
\end{array}
\right).\\
C_{c32}=
R_{y2}(-\pi/4)R_{z12}(-\pi/4)R_{z2}(-\pi/4)R_{y2}(\pi/4) \nonumber \\
=e^{-i\frac{\pi}{4}\hat{e}_{1}\otimes{\hat{\sigma}_{y2}}}e^{-i\frac{\pi}{4}\hat{\sigma}_{z1}\otimes{\hat{\sigma}_{z2}}}e^{-i\frac{\pi}{4}\hat{e}_{1}\otimes{\hat{\sigma}_{z2}}}e^{i\frac{\pi}{4}\hat{e}_{1}\otimes{\hat{\sigma}_{y2}}} \nonumber \\
=\left(
\begin{array}{cccc}
0 & -i & 0 & 0\\
-i & 0 & 0 & 0\\
0 & 0 & 1 & 0\\
0 & 0 & 0 & 1
\end{array}
\right).\\
C_{c41}=
R_{x1}(-\pi/4)R_{z12}(-\pi/4)R_{z1}(-\pi/4)R_{x1}(\pi/4) \nonumber \\
=e^{-i\frac{\pi}{4}{\hat{\sigma}_{x1}}\otimes{\hat{e}_{2}}}e^{-i\frac{\pi}{4}\hat{\sigma}_{z1}\otimes{\hat{\sigma}_{z2}}}e^{-i\frac{\pi}{4}\hat{\sigma}_{z1}\otimes{\hat{e}_{2}}}e^{i\frac{\pi}{4}\hat{\sigma}_{x1}\otimes{\hat{e}_{2}}} \nonumber \\
=\left(
\begin{array}{cccc}
0 & 0 & 1 & 0\\
0 & 1 & 0 & 0\\
-1 & 0 & 0 & 0\\
0 & 0 & 0 & 1
\end{array}
\right).\\
C_{c42}=
R_{x2}(-\pi/4)R_{z12}(-\pi/4)R_{z2}(-\pi/4)R_{x2}(\pi/4) \nonumber \\
=e^{-i\frac{\pi}{4}\hat{e}_{1}\otimes{\hat{\sigma}_{x2}}}e^{-i\frac{\pi}{4}\hat{\sigma}_{z1}\otimes{\hat{\sigma}_{z2}}}e^{-i\frac{\pi}{4}\hat{e}_{1}\otimes{\hat{\sigma}_{z2}}}e^{i\frac{\pi}{4}\hat{e}_{1}\otimes{\hat{\sigma}_{x2}}} \nonumber \\
=\left(
\begin{array}{cccc}
0 & 1 & 0 & 0\\
-1 & 0 & 0 & 0\\
0 & 0 & 1 & 0\\
0 & 0 & 0 & 1
\end{array}
\right).\\
C_{c51}=
R_{y1}(-\pi/4)R_{z12}(-\pi/4)R_{z1}(\pi/4)R_{y1}(\pi/4) \nonumber \\
=e^{-i\frac{\pi}{4}{\hat{\sigma}_{y1}}\otimes{\hat{e}_{2}}}e^{-i\frac{\pi}{4}\hat{\sigma}_{z1}\otimes{\hat{\sigma}_{z2}}}e^{-i\frac{\pi}{4}\hat{\sigma}_{z1}\otimes{\hat{e}_{2}}}e^{i\frac{\pi}{4}\hat{\sigma}_{y1}\otimes{\hat{e}_{2}}} \nonumber \\
=\left(
\begin{array}{cccc}
1 & 0 & 0 & 0\\
0 & 0 & 0 & i\\
0 & 0 & 1 & 0\\
0 & i & 0 & 0
\end{array}
\right).\\
C_{c52}=
R_{y2}(-\pi/4)R_{z12}(-\pi/4)R_{z2}(\pi/4)R_{y2}(\pi/4) \nonumber \\
=e^{-i\frac{\pi}{4}\hat{e}_{1}\otimes{\hat{\sigma}_{y2}}}e^{-i\frac{\pi}{4}\hat{\sigma}_{z1}\otimes{\hat{\sigma}_{z2}}}e^{i\frac{\pi}{4}\hat{e}_{1}\otimes{\hat{\sigma}_{z2}}}e^{i\frac{\pi}{4}\hat{e}_{1}\otimes{\hat{\sigma}_{y2}}} \nonumber \\
=\left(
\begin{array}{cccc}
1 & 0 & 0 & 0\\
0 & 1 & 0 & 0\\
0 & 0 & 0 & i\\
0 & 0 & i & 0
\end{array}
\right).\\
C_{c61}=
R_{x1}(-\pi/4)R_{z12}(-\pi/4)R_{z1}(\pi/4)R_{x1}(\pi/4) \nonumber \\
=e^{-i\frac{\pi}{4}{\hat{\sigma}_{x1}}\otimes{\hat{e}_{2}}}e^{-i\frac{\pi}{4}\hat{\sigma}_{z1}\otimes{\hat{\sigma}_{z2}}}e^{i\frac{\pi}{4}\hat{\sigma}_{z1}\otimes{\hat{e}_{2}}}e^{i\frac{\pi}{4}\hat{\sigma}_{x1}\otimes{\hat{e}_{2}}} \nonumber \\
=\left(
\begin{array}{cccc}
1 & 0 & 0 & 0\\
0 & 0 & 0 & -1\\
0 & 0 & 1 & 0\\
0 & 1 & 0 & 0
\end{array}
\right).\\
C_{c62}=
R_{x2}(-\pi/4)R_{z12}(-\pi/4)R_{z2}(\pi/4)R_{x2}(\pi/4) \nonumber \\
=e^{-i\frac{\pi}{4}\hat{e}_{1}\otimes{\hat{\sigma}_{x2}}}e^{-i\frac{\pi}{4}\hat{\sigma}_{z1}\otimes{\hat{\sigma}_{z2}}}e^{i\frac{\pi}{4}\hat{e}_{1}\otimes{\hat{\sigma}_{z2}}}e^{i\frac{\pi}{4}\hat{e}_{1}\otimes{\hat{\sigma}_{x2}}} \nonumber \\
=\left(
\begin{array}{cccc}
1 & 0 & 0 & 0\\
0 & 1 & 0 & 0\\
0 & 0 & 0 & -1\\
0 & 0 & 1 & 0
\end{array}
\right).\\
C_{c71}=R_{x2}(-\pi/4)R_{z12}(\pi/4)R_{z2}(\pi/4)R_{x2}(\pi/4) \nonumber \\
=e^{-i\frac{\pi}{4}\hat{e}_{1}\otimes{\hat{\sigma}_{x2}}}e^{i\frac{\pi}{4}\hat{\sigma}_{z1}\otimes{\hat{\sigma}_{z2}}}e^{i\frac{\pi}{4}\hat{e}_{1}\otimes{\hat{\sigma}_{z2}}}e^{i\frac{\pi}{4}\hat{e}_{1}\otimes{\hat{\sigma}_{x2}}} \nonumber \\
=\left(
\begin{array}{cccc}
0 &-1 & 0 & 0\\
1 & 0 & 0 & 0\\
0 & 0 & 1 & 0\\
0 & 0 & 0 & 1
\end{array}
\right).\\
C_{c72}=
R_{x1}(-\pi/4)R_{z12}(\pi/4)R_{z1}(\pi/4)R_{x1}(\pi/4) \nonumber \\
=e^{-i\frac{\pi}{4}{\hat{\sigma}_{x1}}\otimes{\hat{e}_{2}}}e^{i\frac{\pi}{4}\hat{\sigma}_{z1}\otimes{\hat{\sigma}_{z2}}}e^{i\frac{\pi}{4}\hat{\sigma}_{z1}\otimes{\hat{e}_{2}}}e^{i\frac{\pi}{4}\hat{\sigma}_{x1}\otimes{\hat{e}_{2}}} \nonumber \\
=\left(
\begin{array}{cccc}
0 & 0 &-1 & 0\\
0 & 1 & 0 & 0\\
1 & 0 & 0 & 0\\
0 & 0 & 0 & 1
\end{array}
\right).\\
C_{c81}=R_{y2}(-\pi/4)R_{z12}(\pi/4)R_{z2}(\pi/4)R_{y2}(\pi/4) \nonumber \\
=e^{-i\frac{\pi}{4}\hat{e}_{1}\otimes{\hat{\sigma}_{y2}}}e^{i\frac{\pi}{4}\hat{\sigma}_{z1}\otimes{\hat{\sigma}_{z2}}}e^{i\frac{\pi}{4}\hat{e}_{1}\otimes{\hat{\sigma}_{z2}}}e^{i\frac{\pi}{4}\hat{e}_{1}\otimes{\hat{\sigma}_{y2}}} \nonumber \\
=\left(
\begin{array}{cccc}
0 & i & 0 & 0\\
i & 0 & 0 & 0\\
0 & 0 & 1 & 0\\
0 & 0 & 0 & 1
\end{array}
\right).\\
C_{c82}=R_{y1}(-\pi/4)R_{z12}(\pi/4)R_{z1}(\pi/4)R_{y1}(\pi/4) \nonumber \\
=e^{-i\frac{\pi}{4}{\hat{\sigma}_{y1}}\otimes{\hat{e}_{2}}}e^{i\frac{\pi}{4}\hat{\sigma}_{z1}\otimes{\hat{\sigma}_{z2}}}e^{i\frac{\pi}{4}\hat{\sigma}_{z1}\otimes{\hat{e}_{2}}}e^{i\frac{\pi}{4}\hat{\sigma}_{y1}\otimes{\hat{e}_{2}}} \nonumber \\
=\left(
\begin{array}{cccc}
0 & 0 & i & 0\\
0 & 1 & 0 & 0\\
i & 0 & 0 & 0\\
0 & 0 & 0 & 1
\end{array}
\right).
\end{eqnarray}
We have obtained 16 CNOT matrices or operators i.e., (4-5,10-23), which are 
obtained by using the similar matrices properties. It means we can perform
the CNOT operation on qubits with the different pulse rotations around the 
different axes according to our convenience. 
\section{Conclusion}
By using the properties of similar matrices we can reduce to some extent the 
physical realization of construction of quantum computer by having different 
number of pulse choices around different axes or different number of CNOT 
matrices. For example, the nuclear magnetic resonance (NMR) apparatus does not 
allow us to send pulse to some concrete direction, if we have one CNOT matrix.
It means we have to modify our apparatus or by new one. But when we have 16 
number of CNOT matrices, we can find compromise with our apparatus. 

Moreover, the technique of similar matrices are very closely related to the 
tensor product, which was observed by us earlier [3]. The similar matrices 
method can be applied to the other branches of quantum theory.

\end{document}